# Finding Influential Bloggers

Bogdan Gliwa and Anna Zygmunt

*Abstract*—Blogging is a popular way of expressing opinions and discussing topics. Bloggers demonstrate different levels of commitment and most interesting are influential bloggers. Around such bloggers, the groups are forming, which concentrate users sharing similar interests. Finding such bloggers is an important task and has many applications e.g. marketing, business, politics. Influential ones affect others which is related to the process of diffusion. However, there is no objective way to telling which blogger is more influential. Therefore, researchers take into consideration different criteria to assess bloggers (e.g. SNA centrality measures). In this paper we propose new, efficient method for influential bloggers discovery which is based on relation of commenting in blogger's thread and is defined on bloggers level. Next, we compare results with other, comparative method proposed by Agarwal *et al.* called iFinder which is based on links between posts.

*Index Terms*—Blogosphere, influential bloggers, social media, social network analysis.

## I. INTRODUCTION

Among nodes in a social network one can identify those that are more important and influential than others in the context of events occurring on the network. Finding such users is an interesting research problem having many practical applications in business, viral marketing, politics, psychology, sociology, etc. But identification influential people is not so easy task. Using basic social network analysis measures such as degree centrality, betweenness centrality or closeness centrality is not enough. People may play different roles in the context of different kind of social media. It is therefore necessary to take into account the characteristics of given social media in determining social influence and roles played by the users. In this paper a comparison of two approaches for finding influential users is presented.

The remaining of this paper is organized as follows: in Section II we present short review, in Section III we provide formal descriptions of these two models. Section IV presents experimental evaluation. In Section V we conclude and present future directions.

## II. RELATED WORK

### A. Blogosphere and Social Network Analysis

Nowadays people use various forms of social media in different ways: to show their status, to be in touch with friends, to express their opinions, to display photos with friends, etc.

Manuscript received October 9, 2014; revised December 12, 2014. This work was supported by the dean grant.
The authors are with AGH University of Science and Technology, Cracow, Poland (e-mail: bgliwa@agh.edu.pl, azygmunt@agh.edu.pl).

Internet social media (e.g. blogs, forums, media sharing systems, microblogging, social networking) has revolutionized the way of communication between people. Blogs play a special role in creating opinions and information propagation. Members demonstrate different activity levels, but generally only a small percentage of them are active.

Structure of blogosphere is as follows. Author gives opinions on some themes or describes interesting events and readers comment on these posts. Posts can be categorized by tags. A very important element of blogs is the possibility of adding comments, which allows discussions. So the basic interactions between bloggers are writing comments in relation to posts or other comments. The relationships between bloggers are very dynamic and temporal. Blogs, posts and comments form a network, which can be analysed by Social Network Analysis (SNA) methods [1]. The SNA approach provides measures (SNA centrality measures) which make it possible to determine the most important or influential nodes (bloggers) in the network. Around such bloggers, the groups are forming, sharing similar interests.

### B. Influential Users in Social Networks

There are many definitions of social roles, which are mainly dependent on the application. For example, from Twitter users one can distinguish *leaders* (who start tweeting, but do not follow others and can have many followers); {*lurkers*} (generally inactive, but occasionally follow some tweets); *spammers* (the unwanted tweeters, also called *twammers*), and *close associates* (including friends, family members, relatives, colleagues, etc.) [2]. In social media, a definition of a role that seems to be most appropriate treats it as a set of characteristics (relevant metrics) that describe behavior of individuals and their interactions between them within a social context [3]. Since it was noticed that most messages were written by a small percentage of users, many early research focused on finding only important (in terms of activity) users. Kaller in [4] noted that active person does not necessarily have to be important and influential in the community. He introduced several characteristics of such influential user, e.g., recognition, generation activity, novelty, eloquence. The most commonly used definition of social influence is that given by Erchul and Raven and cited in [5]: " change in the belief, attitude, or behavior of a person [...], which results from the action, or presence, of another person [...]". Nowadays, identification of influential users is one of the intensely developed directions of research on social networks. Referring to the fundamental article [4] several attempts were made to translate influential users characteristics (e.g. recognition, generation activity, novelty, eloquence) described therein into SNA measures [6].

Influential people affect others which is related to the process of diffusion. Among various centrality measures, *PageRank* seems to be most useful for using in developing algorithms of finding influential users. Many authors try to combine philosophy behind *PageRank* (users activity and connectivity) with another characteristics of influential users. In [7] new measure called *ProfileRank* was introduced. Index similar to h-index (introduced by Hirsch) used to evaluate scientists work was used in [8]. Overview and comparison of several methods and algorithms can be found in [5].

## III. MODEL

In this section we describe comparative method for finding influential bloggers (referenced later as *Agarwal's method*), proposed method for finding such users and, finally, review of methods used to compare rankings of users.

### A. Agarwal's Method for Finding Influential Posts and Bloggers as a Comparative Method

Agarwal *et al.* in [9] proposed a method for finding influential posts and bloggers (called *iFinder*). Their method is based on links between posts (when in the text of a post its author puts url to another post) and they defined *InfluenceFlow* (it is a sum of influence from incoming links (called *inlinks*) minus sum of influence from outgoing links (called *outlinks*)):

$$InfluenceFlow(p) = w_{in}\sum_m I(p_m) - w_{out}\sum_n I(p_n) \quad (1)$$

where $w_{in}$ and $w_{out}$ are weights adjusting contribution of incoming and outgoing influence.

Influence for a post $p$ has the following form:

$$I(p) = w(\lambda) \times (w_c \gamma_c + InfluenceFlow(p)) \quad (2)$$

where $w_c$ is a weight adjusting contribution from number of comments for a post $p$, $\gamma_c$ is a number of comments for a given post and $w(\lambda)$ is a function used to assess eloquence of a post (Agarwal et al. used length of a post as a rough approximation for this feature, but they admitted in their paper that this value it is only useful in comparison posts on different datasets and for comparison on the same datasets is immaterial, so we omitted this measure in our experiments).

The method assume iterative calculation of above described measures until the stable state is reached or fixed number of iterations was performed. In a matrix form the method is described in Fig. 1 (adjacency matrix is created based on links between posts and if a post doesn't have any outlinks then in a row corresponding to that post value of *1/postsNumber* is used to fill that row):

**Algorithm 1:** Agarwal's method for finding influential posts

**Input** : Given a set of blog posts $P$, number of iterations *iter*, similarity threshold $\tau$
**Output**: The influence vector $\vec{i}$ which represents the influence scores of all the blog posts in $P$

```
begin
    Compute the adjacency matrix A between posts;
    Compute vectors λ⃗, γ⃗;
    Initialize i⃗ ← i⃗₀;
    repeat
        i⃗' = λ⃗(w_c γ⃗ + (w_in A^T − w_out A)i⃗);
        iter ← iter − 1;
    until (cos_sim(i⃗, i⃗') < τ) ∨ (iter ≥ 0);
end
```

Fig. 1. Agarwal's algorithm for finding influential posts.

In experiments the influence vector was initialized using value 0.5.

After finding influential posts, Agarwal suggests two strategies for determining influential bloggers – using maximum value of influence for their posts or using average value of influence for their posts. In this paper we used average value of bloggers posts as a metrics describing their influence.

### B. Proposed Method for Finding Influential Bloggers

In paper [10] we introduced roles based on influence. This method of evaluating bloggers influence is based on relation of commenting in blogger's thread (so the link in a network from blogger $A$ to $B$ is when $A$ commented in some of $B$' posts and both $A$ and $B$ have to be bloggers i.e. they have to write posts, not only comments). Furthermore, the method is defined on bloggers level (so we determine influence strictly for authors and not for posts, as it was performed in Agarwal's method).

We defined measure *PostInfluence* (denoted *PInf* assessing quality of written posts by bloggers) which was extended in this study and had the following form (this measure is defined for an author $a$):

$$PInf(a) = \sum_{p_a} f(pr(p_a)) + w\sum_v \frac{c(v,a)}{\sum_k c(v,k)} PInf(v) \quad (3)$$

where $p_a$ means a post of blogger $a$, $pr(p_a)$ is a function determining number of comments for a post $p_a$ excluding the author's comments in his own thread (called *post response*), $f$ is a function that assigns scores for a given post response, $v$ means a blogger who commented some of $a$' posts, $c(v,a)$ denotes number of comments of blogger $v$ in posts of $a$, $k$ means a blogger who was commented by blogger $v$ and $w$ is a weight adjusting contribution of the second part in evaluation of influence.

Formula $\frac{c(v,a)}{\sum_k c(v,k)}$ describes proportion of comments from blogger $v$ to blogger $a$ in comparison with all comments of $v$. Adjacency matrix is constructed using this formula (moreover, if a blogger $a$ gave no comments to others then in that matrix for a row corresponding to that user we fill values equal to *1/bloggersNumber* – such procedure is related with stability of iterative process).

As we can see, *PostInfluence* has two parts: first one which assesses response from other users (not only bloggers, also from people who only comments, but excluding author's own comments) and second one which describes position of a blogger in a network (based on the fact of commenting by other bloggers which is a form of recognition in a community of bloggers). The first part in a matrix notation is marked as $\beta$. The second part is similar to *PageRank* formula when we consider weighted network. In a matrix form the method is depicted in Fig. 2.

As a function determining scores for post responses we used the following one:

$$f(x) = \begin{cases} -0.1, & \text{when } x \leq 1 \\ e^{-\lambda_h(m_h-x)}, & \text{when } x > 0.25 * maxResp \\ e^{-\lambda_l(m_l-x)}, & \text{otherwise} \end{cases} \quad (4)$$

where *maxResponse* is maximum value of post response in a given time (based on the values achieved for bloggers).

**Algorithm 2:** Proposed method for finding influential authors
**Input** : Given a set of blog posts $P$ and set of bloggers $B$, number of iterations *iter*, similarity threshold $\tau$
**Output**: The influence vector $\vec{i}$ which represents the influence scores of all the bloggers in $B$

begin
  Compute the adjacency matrix $A$ between bloggers;
  Compute vector $\vec{\beta}$;
  Initialize $\vec{i} \leftarrow \vec{i_0}$;
  repeat
    $\vec{i'} = \vec{\beta} + wA^T\vec{i}$;
    $iter \leftarrow iter - 1$;
  until $(cos\_sim(\vec{i}, \vec{i'}) < \tau) \vee (iter \geq 0)$;
end

Fig. 2. Proposed algorithm for finding influential bloggers.

First part of this function is a penalty for writing weak posts (with no response). The second and the third part is the same function but with different parameters. Example of such a function is below (see Fig. 3):

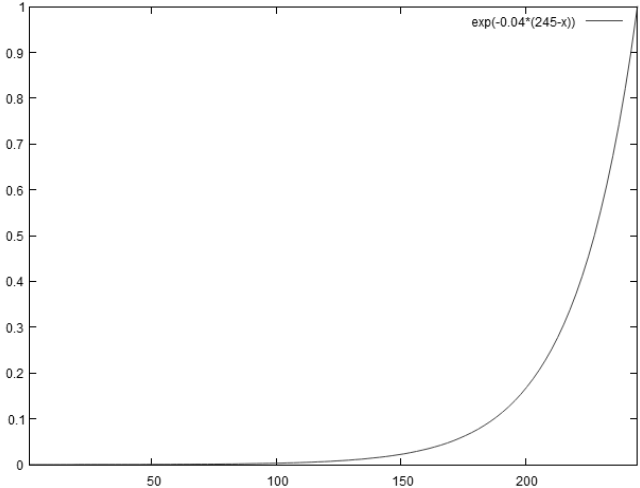

Fig. 3. Example of exponential function used in our formula.

This is exponential function and we think it is better than linear because it is harder to achieve higher numbers of comments for posts. The reason for introduction two parts with that function is that this function fast decreases and the parameters $\lambda_h$, $m_h$, $\lambda_l$ and $m_l$ are adjusted to achieve that the result of the function $f$ for an argument equal to *maxResponse/4* is 0.2 (for *maxResponse* is 1.0) and for 0 is 0.001.

### C. Comparison of Rankings

Let $S$ and $T$ be two rankings (ordered list of elements) and assume that number of elements in those rankings is $N$. Then we can define **overlap** (intersection of two lists) as:

$$Overlap(S,T) = |S \cap T| \quad (5)$$

To define **average overlap AO** and **rank-biased overlap** [11] we need to introduce some additional formulas. Denotation $S_{k:m}$ means sublist of S (from $k$ index to $m$). Overlap of lists $S$ and $T$ to depth $d$ has form:

$$O(S,T,d) = |S_{1:d} \cap T_{1:d}| \quad (6)$$

The overlap of $S$ and $T$ divided by the depth $d$ is called *proportion* [11]:

$$A(S,T,d) = \frac{O(S,T,d)}{d} \quad (7)$$

Using above equations we can define **average overlap** as:

$$AO(S,T) = \frac{1}{N}\sum_{d=1}^{N} A(S,T,d) \quad (8)$$

and **rank-biased overlap**:

$$RBO(S,T,p) = (1-p)\sum_{d=1}^{N} p^{d-1} A(S,T,d) \quad (9)$$

In experiments we used $p$ parameter equals to 0.85.
In rank-biased overlap the agreement between two rankings on top positions is more important than on the next positions (equation (9) has decreasing weights).

## IV. RESULTS

### A. Dataset Description

Experiments were conducted on data from portal Salon24[1] (Polish blogosphere), which contains mostly political discussions and discussions about current events, but also other topics appear. The dataset contains data from time range 1.01.2008-6.07.2013. The whole period of time was divided into disjoint timeslots, each lasting 1 month. As a result 67 slots were formed (last slot was partial so we rejected it).

### B. Inlinks and Outlinks in Posts

In Table I we can observe statistics for inlinks and outlinks from posts. One can see that less than 20% of all outlinks have their targets in the portal. For all such outlinks significant number of them can be matched for an author who is referenced. Situation is much worse when we want to find target post of a reference (26% of urls can be matched to a post from those that can be matched to an author). It is related with the fact that some bloggers use links to main page of other bloggers (instead of particular blog). Moreover, on such blog portals from time to time there is reorganization of content, so urls changes (but links in the content of posts rather not) and some of them which initially were correct, in some time went stale.

TABLE I: STATISTICS FOR INLINKS AND OUTLINKS

| Type of links | Number of links |
|---|---|
| All outlinks | 515 066 |
| All salon24 outlinks | 98 467 (19% of all outlinks) |
| Matched author inlinks | 95 585 |
| Matched post inlinks | 24 802 (26% of all author inlinks) |
| Same author in matched post inlinks | 13 548 (55% of all matched post inlinks) |

In Fig. 4 the number of matched post inlinks in time is presented. In the first slots the number of such links is very small and it may be connected with the start of crawling data (people in that moment referred to data that was not crawled

---
[1] www.salon24.pl

in the dataset).

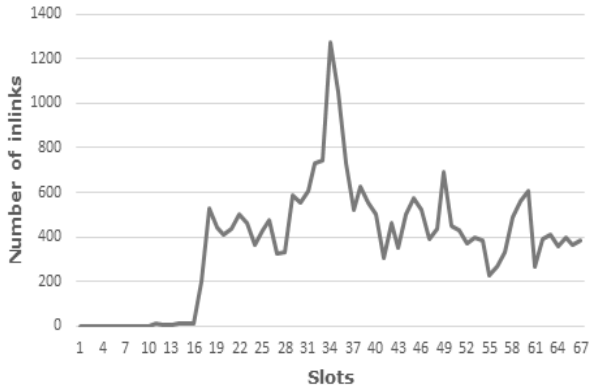

Fig. 4. Number of matched post inlinks in time.

Interesting thing can be observed in Table II, people frequently link to their own posts. In most linked authors there is only one exception.

TABLE II: MOST LINKED AUTHORS FOR THEIR POSTS

| Author | Inlinks | Self inlinks |
|---|---|---|
| KACPRO | 1 321 | 1313 (99%) |
| PANNA WODZIANNA | 1 015 | 945 (93%) |
| GPS.1965 | 944 | 918 (97%) |
| KRZYSZTOF LESKI | 707 | 532 (75%) |
| KRZYSZTOF WOŁODŹKO… | 542 | 519 (96%) |
| BANG BANG | 506 | 496 (98%) |
| MAREK MOJSIEWICZ | 343 | 319 (93%) |
| MAREK MIGALSKI | 339 | 10 (3%) |
| NIEGDYSIEJSZY BLONDYN | 330 | 286 (87%) |
| CARCAJOU | 325 | 237 (73%) |

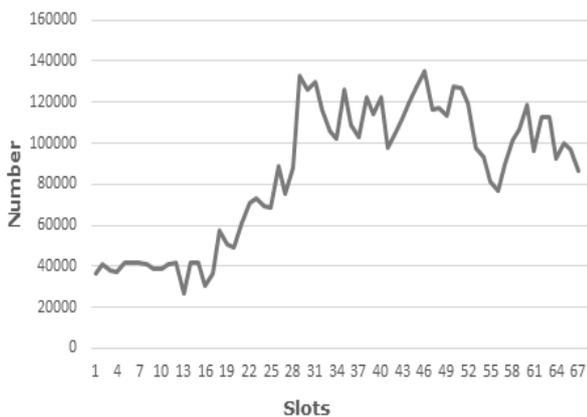

Fig. 5. Number of comments in time.

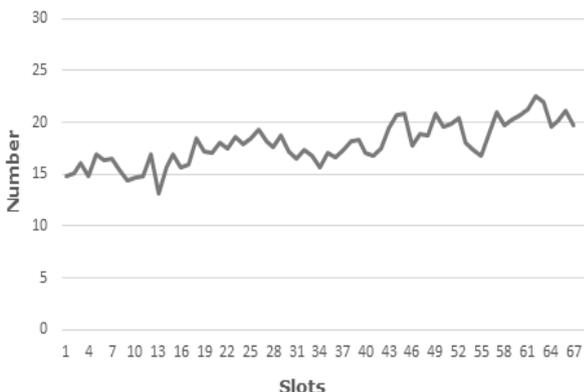

Fig. 6. Average number of comments per post in time.

## C. Comments in Posts

In Fig. 5 we can observe number of comments in time which increased a lot after initial period of time.

Fig. 6 shows ratio of comments to posts. We can see the it is slowly increasing.

Table III presents most comented authors and we can observe that in this case the number of self-referencing is much lower than in the case of links.

TABLE III: MOST COMMENTED AUTHORS

| Author | Comments | Self comments |
|---|---|---|
| FREE YOUR MIND | 151 949 | 52 118 (34%) |
| RENATA RUDECKA… | 113 118 | 48 970 (43%) |
| KRZYSZTOF LESKI | 105 469 | 35 056 (33%) |
| CEZARY KRYSZTOPA | 84 684 | 36 422 (43%) |
| SOWINIEC | 76 099 | 35 456 (47%) |
| STARY | 68 612 | 25 596 (37%) |
| GRZEGORZ WSZOŁEK | 67 991 | 19 849 (29%) |
| UFKA | 67 956 | 16 777 (25%) |
| MAREK MIGALSKI | 61 648 | 1 639 (3%) |
| 1MAUD | 58 575 | 20 307 (35%) |

## D. Comparison of Results between Proposed Method and Aggarwal'S Method

Fig. 7 depicts ratio of posts to bloggers in time. One can see that it is quite stable and around 5. This is very important because proposed method relies on the number of bloggers (matrix of bloggers is present in that method) and Agarwal's method on the number of posts. This indicates clearly that proposed method is much more efficient.

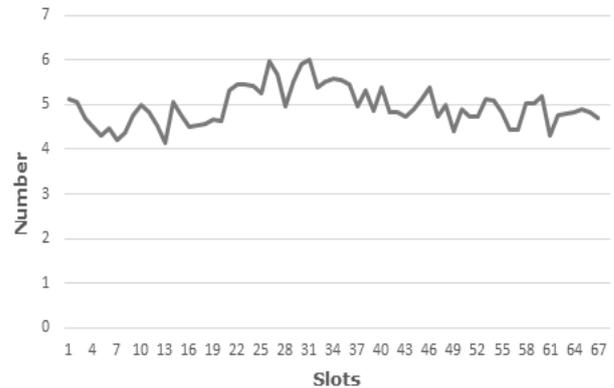

Fig. 7. Number of posts divided by number of bloggers in time.

Table IV and V contain results of both methods – list of bloggers that were most frequent in top 15 in time slots. Table VI summarizes them and we can see that overlap is significant, especially for those top ranked bloggers (*RankBiased overlap* has the biggest value).

TABLE IV: BLOGGERS MOST FREQUENT IN TOP 15 – AGARWAL'S METHOD

| Author | Times |
|---|---|
| FREE YOUR MIND | 55 |
| ŁUKASZ WARZECHA | 53 |
| IGOR JANKE | 37 |
| RENATA RUDECKA… | 37 |
| WOJCIECH SADURSKI | 36 |
| KATARYNA | 31 |
| JAN OSIECKI | 23 |
| KRZYSZTOF KŁOPOTOWSKI | 22 |
| ALEKSANDER ŚCIOS | 20 |
| KATRINE | 20 |

TABLE V: BLOGGERS MOST FREQUENT IN TOP 15 – PROPOSED METHOD

| Author | Times |
|---|---|
| FREE YOUR MIND | 58 |
| IGOR JANKE | 44 |
| ŁUKASZ WARZECHA | 41 |
| RENATA RUDECKA… | 40 |
| WOJCIECH SADURSKI | 35 |
| RYBITZKY | 33 |
| KRZYSZTOF LESKI | 30 |
| TOMASZ TERLIKOWSKI | 29 |
| GRZEGORZ WSZOŁEK | 28 |
| MAREK MIGALSKI | 28 |

TABLE VI: METRICS OF RANK COMPARISON FOR MOST FREQUENT BLOGGERS IN TOP 15 IN BOTH METHODS

| Metrics | Value |
|---|---|
| Overlap | 0.6 |
| Average overlap | 0.726 |
| RankBiased overlap | 0.728 |

In Fig. 8 one can observe similarity of rankings for both methods in time. Generally overlap achieves the highest values (max value is 0.8) and rank-biased overlap the smallest.

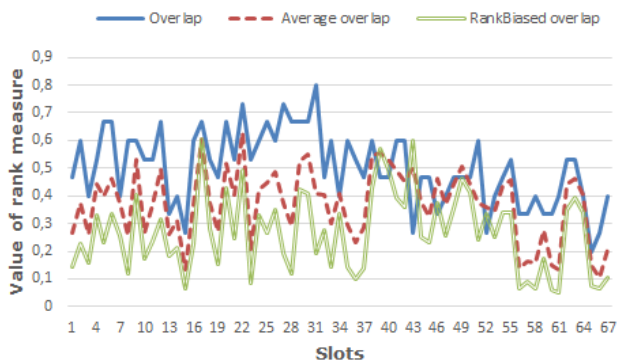

Fig. 8. Comparison of rankings between methods in time.

## V. CONCLUSION AND FUTURE DIRECTIONS

In this paper new method of assessing bloggers influence was introduced. It has several advantages over the comparative method. Firstly, the method is based on the response of other bloggers in the form of a comment which seems more reasonable in this dataset as self-referencing is not as frequent as in links (also number of comments is two orders of magnitude bigger than for links). Secondly, the proposed method is much more effective computationally (typically about 5 times smaller number of rows and columns in matrix to calculate). Moreover, the proposed method relies on the nonlinear function to assess response which seems to be reasonable choice (it is much harder achieve better result in higher values in comparison to lower ones). It is also worth mentioning that proposed method does not count activity of post author in his own thread (e.g. in the response for post the comments of an author of that post are not taken into consideration).

Future works will concern the analysis on other datasets (especially, on datasets from different countries) and detailed study on choice of function to assess scores for response on posts. Another direction of future research is the comparison with PageRank or other methods that can be used to find influential bloggers. Finally, we also plan to study in a more detail links that can be matched to other authors (not to other posts as we did in this study) and compare our results with those achieved using that relation.

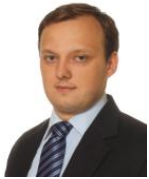

**Bogdan Gliwa** is a PhD student at the Department of Computer Science AGH University Science and Technology, Kraków, Poland. His areas of interest include complex network analysis, data mining and machine learning.

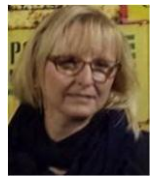

**Anna Zygmunt** is an assistant professor at the Department of Computer Science AGH University of Science and Technology, Kraków, Poland. Her areas of interests include database systems, decision support systems, data mining and social network analysis. She has authored over 80 research articles.